\documentclass[%
 reprint,
superscriptaddress,
 amsmath,amssymb,
 aps,
 prl
]{revtex4-1}

\usepackage{graphicx}
\usepackage{dcolumn}
\usepackage{bm}
\usepackage{xcolor}

\begin{document}

\preprint{APS/123-QED}

\title{Static adhesion hysteresis in elastic structures}

\author{Edvin Memet}
\affiliation{Department of Physics, Harvard University, Cambridge, Massachusetts 02138, USA}
\author{Feodor Hilitski}
\affiliation{Department of Physics, Brandeis University, Waltham, Massachusetts 02453, USA}
\author{Zvonimir Dogic}
\affiliation{Department of Physics, Brandeis University, Waltham, Massachusetts 02453, USA}
\affiliation{Department of Physics, University of California, Santa Barbara, California 93106, USA}
\author{L. Mahadevan}
\email{lmahadev@g.harvard.edu}
\affiliation{Paulson School of Engineering and Applied Sciences, Department of Physics, Department of\\ Organismic and Evolutionary Biology, Harvard University, Cambridge, Massachusetts 02138, USA \looseness=-1}

\date{\today}

\begin{abstract}

Adhesive interactions between elastic structures such as graphene sheets, carbon nanotubes, and microtubules have been shown to exhibit hysteresis due to irrecoverable energy loss associated with bond breakage, even in static (rate-independent) experiments. To understand this phenomenon, we start with a minimal theory for the peeling of a thin sheet from a substrate, coupling the local event of bond breaking to the nonlocal  elastic relaxation of the sheet and show that this can drive static adhesion hysteresis over a bonding/debonding cycle. Using this model we quantify hysteresis in terms of the adhesion and elasticity parameters of the system. This allows us to derive a scaling relation that preserves hysteresis at different levels of granularity while resolving a seeming paradox of lattice trapping in the continuum limit of a discrete fracture process. Finally, to verify our theory, we use new experiments to demonstrate and measure adhesion hysteresis in bundled microtubules.

\end{abstract}

\pacs{Valid PACS appear here}

\maketitle



\section{Introduction}

The ubiquity of hysteretic behavior in peeling, fracture, or adhesion processes  has long been known  in systems spanning many orders of magnitude including graphene and carbon nanotubes \cite{miskin2017, sasaki2008}, gecko adhesion, actin bundling and dissolution \cite{hosek2004}, DNA melting and denaturation \cite{whitelam2008, smith1996,rouzina2001,peyrard2004}, adhering vesicles \cite{dembo1988}, partially frayed dynamic axonemes \cite{aoyama2005}, extensile microtubule bundles that generate autonomous flows \cite{sanchez2012spontaneous}, and elastic contact in soft materials and structures \cite{ maugis2013,pesika2007,molinari2008,cohen2018,williams2014}. Although it has been nearly a century since Obreimoff measured the energy required to split a multilayer mica sheet \cite{obreimoff1930,kendall2007,pocius2002}, and interpreted it in terms of an adhesion energy, the microscopic mechanisms behind hysteresis often remain poorly understood \cite{miskin2017}. While hysteresis is often attributed to velocity-dependent processes \cite{liu2018,chung2005,kendall2007,de2019,villey2015}, numerous observations of static hysteresis have been reported\cite{chen1991,sekiguchi2014,puglisi2013,maddalena2009} such as in the peeling of a thin graphene sheet from a substrate \cite{miskin2017}. Accordingly, theoretical frameworks for static hysteresis have been developed in the context of membrane adhesion \cite{maddalena2009, evans1985a, evans1985b}, lattice trapping \cite{thomson1971, gao1989}, Griffith cracks \cite{rice1978}, adhesive contact \cite{guduru2007, noderer2007, kesari2011}, and composite materials \cite{xia2013, kendall1975}.

In this paper we develop a general theoretical framework for rate-independent adhesion hysteresis in elastic structures. In particular, our model is grounded in experimental observations of this phenomenon in two specific systems with distinct geometries: (i) old experiments involving the peeling of a graphene sheet from a fixed, flat substrate (Fig. 1A, top left) \cite{miskin2017} and (ii) new experimental measurement of hysteresis in the buckling-induced fraying of a pair of bundled microtubules, in which one of the microtubules acts as a curved substrate with a variable shape as a function of strain (Fig. 2A).

\begin{figure*}[ht]
\centering
\includegraphics[width=0.8\textwidth]{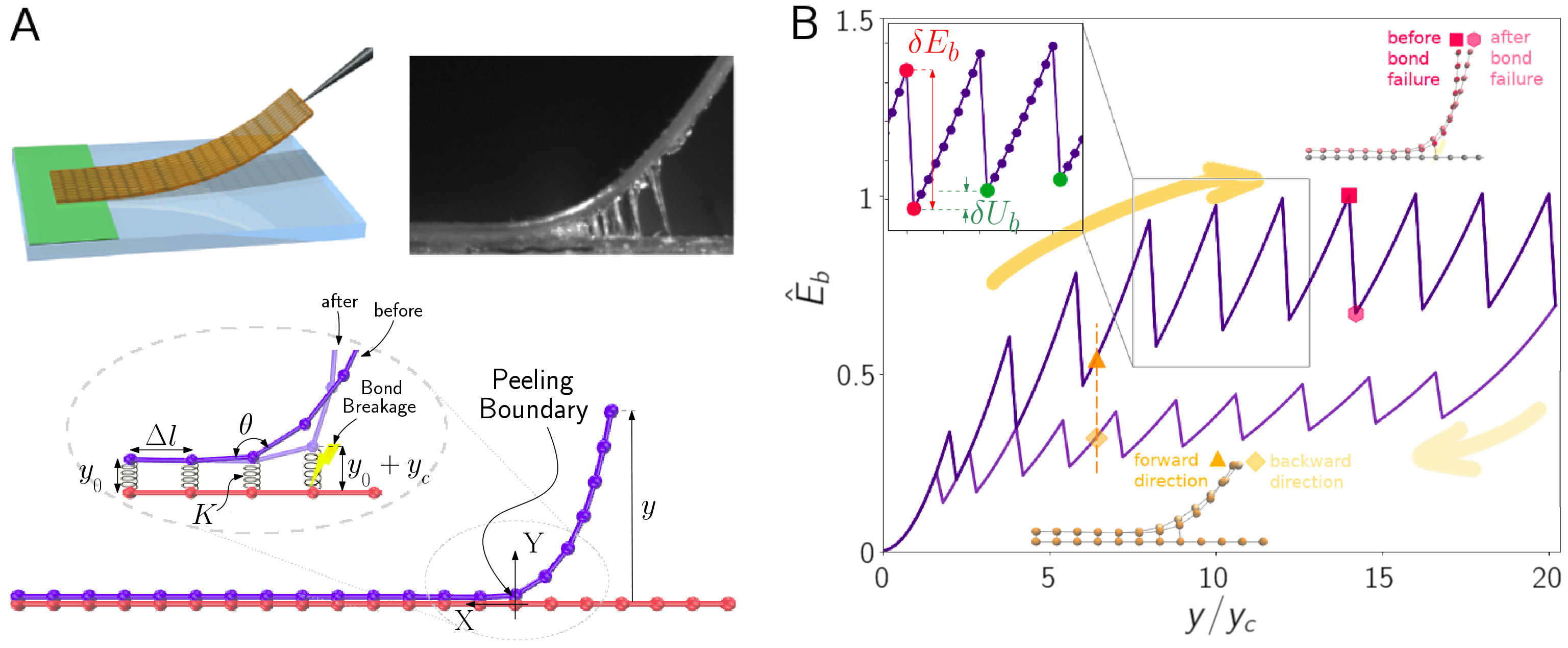}
\caption{(A) Top: Representation of peeling graphene sheet (left) and experimental image of peeling a pressure sensitive adhesive (right). Images reproduced from \cite{miskin2017} and \cite{villey2015}, respectively. Bottom: Discrete elastic chain peeling away from a flat adherent substrate. Zoomed in region illustrates chain and bond rearrangement after bond breakage: as the rightmost bond breaks and moves away from the substrate, remaining bonds stretch more to accomodate the increased stress. (B) Plot of scaled bending energy $\hat E_b = E_b / E_b^0$  versus scaled endpoint displacement $y/y_c$ from simulation ($E_b^0 = B \kappa_c^2 l_H = B y_c ^2 / l_H^3$ is the natural scale for bending energy). Arrows indicate the sequence of motion of the free end displacement: first increasing (upper part), then decreasing back to zero (lower part). Filament configurations are represented visually at pairs of points indicated by red and orange symbols, respectively. (Inset): Sawtooth pattern accompanies bond breakage or re-forming. $\delta E_b$ (red) is the energy loss following single bond breakage, while $\delta U_b$ (green) is the net bending energy change that accompanies peeling of one segment.}
\end{figure*}

\section{Hysteresis in peeling off a flat surface}

\subsection{Equations of motion}

We start by considering two elastic chains interacting with each other adhesively through reversibly breakable, non-hysteretic springs (Fig. 1A). Each chain has $n$ particles spaced apart by $\Delta l \approx L/n$, where $L$ is the length of a chain. The adhesive interaction is associated with breakable elastic links of stiffness $K$, rest length $y_0$, and cutoff $y_0+y_c$ that connect corresponding particles on the two chains. One chain is fixed, acting as a rigid foundation, while the other one initially starts in equilibrium and is quasi-statically loaded and unloaded at one end. The potential energy of such a system is 

\begin{align}
 \Phi = &\frac{1}{2} \left[ \sum_{i=2}^{n-1}  \frac{B}{\Delta l} \left( \theta_{i} - \pi\right)^2 + \sum_{k=1}^{n-1}  k \left(|\boldsymbol{r_{k+1}} - \boldsymbol{r_k}| - \Delta l \right)^2 \right] \nonumber  \\
 +& \frac{1}{2} \sum_{k=1}^{n} \text{min} \left(K \left(|\boldsymbol{r'_k} - \boldsymbol{r_k}|\right)^2, K y_c^2 \right),
\end{align}

\noindent where the first term represents filament bending energy defined in terms of the angle $\theta_i$ formed by triplets of neighboring particles $(i-1, \, i, \, i+1)$ along the mobile chain, the second term corresponds to filament stretching, where $\boldsymbol{r_k} = \left(x_i, y_i\right)$, $\boldsymbol{r'_k} = \left(x_j, y_j\right)$ are position vectors for the mobile and fixed chains, respectively, and $k$ is the intrachain stifness, while the third term -- modeling interfilament adhesion -- corresponds to stretching the links between chains. In the limit of thin  filaments or sheets, the geometric scale separation implies that stretching is very expensive relative to bending (i.e. the material is effectively inextensible), so that we may take the springs connecting particles on the same chain to have a large stiffness, i.e.  $k (\Delta l)^2/B \rightarrow \infty$.

Starting from the energy (1), we can write the overdamped equations of motion for the system as $- \, \mathrm{d} \Phi / \mathrm{d}  \boldsymbol{r_k} = \gamma_t \dot{\boldsymbol{r_k}}$ and $-\, \mathrm{d} \Phi / \mathrm{d} \theta_i = \gamma_r \dot \theta_i$, where $\gamma_t$ and $\gamma_r$ are translational and rotational damping coefficients. We note that a similar model was used by Thomson to study the lattice trapping of fracture cracks ~\cite{thomson1971}. However, Thomson's classic theory only works in the discrete limit, with the trapping effect vanishing in the continuum limit. As we will show, our framework provides a self-consistent way of taking the continuum limit while still having lattice trapping/hysteresis.

\subsection{Length scales and dimensionless parameters}

Our system is characterized by three independent length scales: a lattice (discrete) length scale $\Delta l$, a maximum displacement associated with adhesive bond breakage $y_c$, and the radius of curvature at the peeling boundary 1/$\kappa_c$. Geometrically, the critical curvature can be expressed in terms of the ``rise'' $y_c$ and ``run'' $l_H$: $\kappa_c \sim y_c /l_H^2$. The latter, $l_H$, is called the healing length  and determines how far the perturbation effectively extends into the bonded region (Eq. 3). That is, to the right of the peeling boundary (Fig. 1A) we have the perturbed (peeled) domain, while far enough inside the bulk, to the left of the boundary, the perturbation decays exponentially; $l_H$ sets the scale of the transition zone between the perturbed domain and the unperturbed bulk region \citep{zapperi2011,janosi1998}. These three independent length scales generate two independent dimensionless quantities (besides $n$): $\Delta l/l_H$, which characterizes the  mechanical response along the filament direction, and $y_c/l_H$, which relates to the peeling angle (see SM).

\subsection{Continuum theory}

A continuum theory for the height profile $y \left(x\right)$ inside the bonded region $x>0$ (Fig. 1A and Eq. 2) in the limit of small slopes and deformations provides a quantitative value of the healing length $l_H$. Indeed, by coarse-graining the discrete energy (1) over length scales large compared to the spacing $\Delta l$ between bonds, replacing differences by derivatives ($\pi - \theta \rightarrow y'; \, \boldsymbol{r'_k} - \boldsymbol{r_k} \rightarrow y\left(x\right) $), we find that the Euler-Lagrange equation associated with the continuum version of the functional (1) is given by \citep{zapperi2011}
\begin{equation}
B y'''' + \frac{K}{\Delta l} \left(y- y_0\right) = 0.
\end{equation}

\begin{figure*}[ht]
\includegraphics[width=\textwidth]{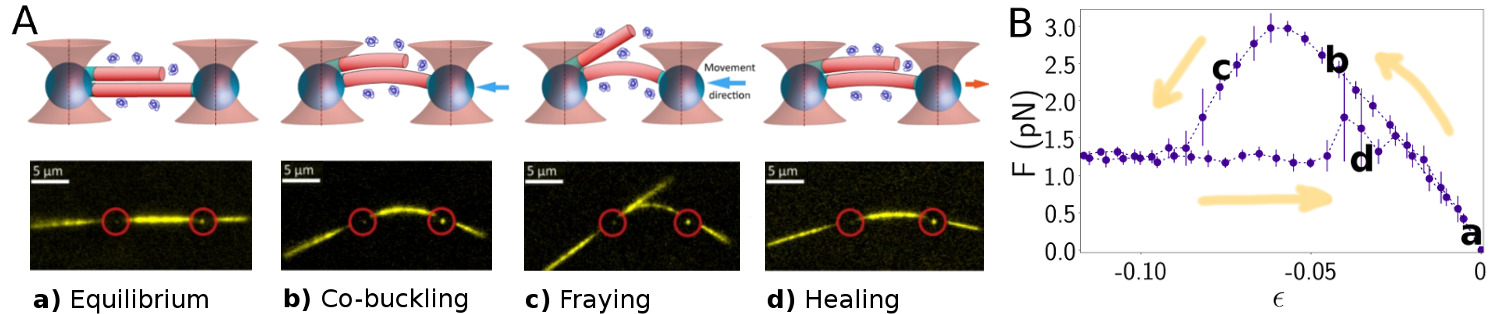}
\caption{(A) Fraying of a composite MT bundle in response to a tensile force applied with optical tweezers (bottom) alongside corresponding schematics (top, not to-scale). Red circles indicate trap positions.  (B)  Measured force-strain exhibited hysteresis associated with bundle fraying and rehealing. Filament configurations corresponding to points (a) to (d) are shown in panel A. Arrows indicate the measured force as the optical trap applies buckling forces and subsequently relaxes back towards the equilibrium. Strain $\epsilon$ is defined as $\epsilon = (d - L)/L$, where $d$ is the bead separation and $L$ is the filament length between the two attachment points.}
\end{figure*}

\noindent With boundary conditions  $y\left( \infty \right) \rightarrow y_0, \,\, y'\left( \infty \right) \rightarrow 0, $, and $y\left( 0 \right) = y_0 + y_c, \,\, B y'''\left( 0 \right) = F$ (the vertical applied peeling force), the solutions for $y\left(x\right)$  and $\kappa\left(x \right)=y''(x)$ are

\begin{align} 
y \left(x\right) &= y_0 + y_c \, e^{-x/l_H} \cos\left(x/l_H\right), \\
\displaystyle \kappa \left(x\right) &= {\frac{2 y_c}{l_H^2}} \, e^{-x/l_H} \sin\left(x/l_H\right),
\end{align}

\noindent where the healing length 
\begin{equation}
l_H = \sqrt{2} \left( \frac{B \Delta l}{K}\right)^{1/4}.
\end{equation}
and the peeling force $F=2 y_c/Bl_H^3$. From Eq. (4), the curvature at the peeling boundary is $\kappa_c = \kappa\left(0\right)  = 2  y_c/l_H^2$ (as predicted geometrically in terms of the ``rise'' and ``run''), which can be rewritten in a more familiar form \citep{stewart1987,majidi2012} in terms of the adhesion energy per unit length $J = n K y_c^2 / (2L)$: 

\begin{equation}
\kappa_c = \kappa\left(0\right) = \frac{2 y_c}{l_H^2} = \sqrt{\frac{2J}{B}}
\end{equation}

\subsection{Simulations}

To compare our results of the simple continuum model with those obtained from our discrete model for the energy given by (Eq. 1), we simulate the dynamics of the adhesive interaction via an overdamped viscous relaxation numerical method~\cite{Arnold2013} (see SM for details). We find that, as we quasi-statically raise one end of the mobile chain, corresponding to the loading phase, bonded segments successively peel from the substrate, as the boundary that separates the bonded and debonded phases advances (Fig. 1A). In the unloading phase, we reverse the displacement direction of the free end, which causes debonded segments to successively re-enter the interaction range and thus re-adhere to the substrate, leading to healing. The healing pathway is mechanically and thermodynamically different from the peeling pathway, a hallmark of hysteresis. Hysteresis is apparent, for example, in a plot of scaled bending energy versus strain (Fig. 1B). The same plot also reveals a characteristic pinning-depinning ``sawtooth'' pattern, which arises from alternating cycles of bending energy accumulation and sudden bond breakage \citep{maddalena2009,puglisi2013}. 

\begin{figure*}[ht]
\includegraphics[width=0.95\textwidth]{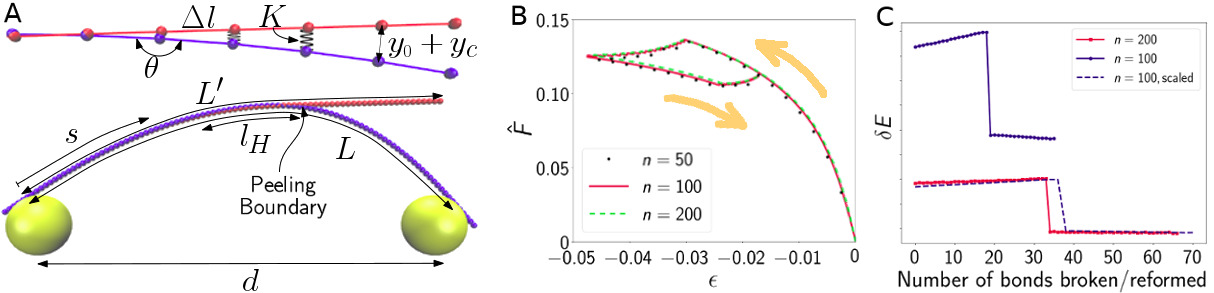}
\caption{(A) Generalized elastic chain model in which both filaments are free to deform. The longer filament (blue) is attached rigidly to two optical beads (yellow), while the shorter, adhering filament (red) is attached to only one bead. The right optical bead is mobile, while the left one is fixed.  (B) Scaled force $\hat F = F/\left(B/L^2\right)$ plotted as a function of imposed strain $\epsilon$ for three simulations with parameter scaling as $K \rightarrow n^3 K$, $y_c \rightarrow y_c/n^2$. (C) Energy loss per bond broken or re-formed versus the cumulative number of bonds broken/re-formed for two of the simulations in panel B, with n = 100 (blue) and n = 200 (red). The two profiles can be made to collapse (signifying equal scaled hysteresis size) by scaling the horizontal axis by a factor of $\Delta l$ and the vertical axis by the bending
energy scale $E_b^0$.}
\end{figure*}

\subsection{Quantification of hysteresis}

When a bond breaks, stress redistribution causes it (and the rest of the free chain) to move further away from the range of the adhesive potential (Fig. 1A, inset) such that on the way back it needs to travel more in order to re-form. Thus, even if individual bonds are not intrinsically hysteretic, macroscopic hysteresis will still emerge via the coupling of a local event (bond breaking) to a nonlocal event (overall elastic relaxation). Indeed, simulations show that the decrease in bending energy $\delta E_b$ upon bond breakage (Fig. 1B, inset) is only partially balanced by an increased load on the remaining springs, i.e. an increase  in the adhesion energy $\delta E_s$. This imbalance results in a net energy loss for the filament $\delta E = \delta E_b + \delta E_s <0$. Meanwhile, when a bond reforms, $\delta E_b >0$ and $\delta E_s <0 $ such that we still have net energy dissipation $\delta E < 0$. 

We expect bending and adhesion energy jumps ($\delta E_b$ and $\delta E_s$) to scale as the energy of a single bond $K y_c^2$: $\delta E_b, \, \delta E_s \sim K y_c^2$. This can also be written in terms of the natural bending energy scale $E_b^0$ of the entire filament, $E_b^0 \equiv B \kappa_c^2 l_H$ (since $l_H$ rather than $L$ is the scale of the deformed region): 

\begin{equation}
\delta E_b, \, \delta E_s \sim K y_c^2 \sim E_b^0 \, \frac{\Delta l}{l_H}
\end{equation}

\noindent The same scaling also applies to $\delta U_b$, the bending energy change across a single pinning-depinning cycle (Fig. 1B, inset). Indeed, since the end result of the cyle is the peeling of a single segment of size $\Delta l$, we might expect $\delta U_b \sim B \kappa_c^2 \Delta l \sim B \kappa_c^2 l_H \times  \left(\Delta l/l_H\right) \sim  E_b^0 \Delta l/l_H$. 

However, it should be apparent that the net energy loss $\delta E = \delta E_b + \delta E_s$ should scale differently from $\delta E_b$ and $\delta E_s$. For example, as $\Delta l/l_H \rightarrow 0$ we expect the breaking of a single bond to have a negligible effect on the shape of the peeled filament and on the stress distribution, meaning that we must have $\delta E \rightarrow 0$ as $\Delta l/l_H \rightarrow 0$. The simplest scaling satisfying this requirement is $\delta E \sim K y_c^2 \times \left(\Delta l/l_H\right) \sim E_b^0 \times \left(\Delta l/l_H\right)^2$, which is confirmed by simulations (Fig. S2B). In order to obtain the \textit{dimensionless} energy loss due to the breakage of a single bond we divide $\delta E$ by the bending energy scale $E_b^0$: 

\begin{equation}
\delta e = \delta E / E_b^0 \sim \left(\Delta l /l_H\right)^2. 
\end{equation}

\subsection{Parameter scalings that preserve hysteresis}

As we change the discretization $n$, hysteresis size will change if we naively scale the spring stiffness $K$ inversely with $n$ (Fig. S4). Therefore, in order to find the scaling that will render hysteresis independent of $n$, we require the invariance of several quantities: (i) the dimensionless energy loss summed across all segments, which we approximate as $n \delta e$, (ii) the adhesion energy per unit length $J = n K y_c^2 /(2L)$, and (iii) the curvature at the scaling boundary $\kappa_c$ (Eq. 6). Substituting $K = 2 J L / \left(n y_c^2 \right)$ and $l_H$ (Eq. 5) in the expression for $\delta e$ (Eq. 8) we get

\begin{equation}
n \, \delta e \sim \frac{\kappa_c L^2}{n\, y_c}.
\end{equation}

\noindent Considering $L$ to be fixed, invariance of the energy loss $n \delta e$ in the continuum limit implies $y_c \sim 1/n $, while invariance of $J$ then requires $K \sim n$. As a result, $l_H \sim (n K)^{-1/4} \sim 1/\sqrt{n}$, which means that in the continuum limit  the transition between the bonded and debonded regions occurs instantly ($l_H \rightarrow 0$), without a weakly bound intermediate region. 

\subsection{Graphene peeling experiments}

To test our theory on real data, we start with observations\cite{miskin2017} of substantial hysteresis in peeling a graphene sheet from a flat surface, with the energy required for delamination reported to be 100 times larger than the energy recovered upon readhesion, but with no explanation given. Using the experimental parameters for the graphene sheet of length $L = 60\, \mu\mathrm{m}$, bending rigidity $B \approx 3 \times 10^9 \, pN \times \mu \mathrm{m}^2$, and effective adhesion energy per unit length $J \approx 10^6 \, pN$, and choosing a discretization size $n = 80$, our simulations yield that there is a factor of $\sim 10$ difference between curvature $\kappa_c$ in the peeling regime compared to that in the healing regime (Fig. S6B) seen in experiments, and a peeling front displacement of around 5 $\mu\mathrm{m}$ per $\mu \mathrm{m}$ of vertical displacement (Fig. S6A, inset), which is also close (within a factor of two) to the measured value\cite{miskin2017}.

\begin{figure}[ht]
\includegraphics[width=\columnwidth]{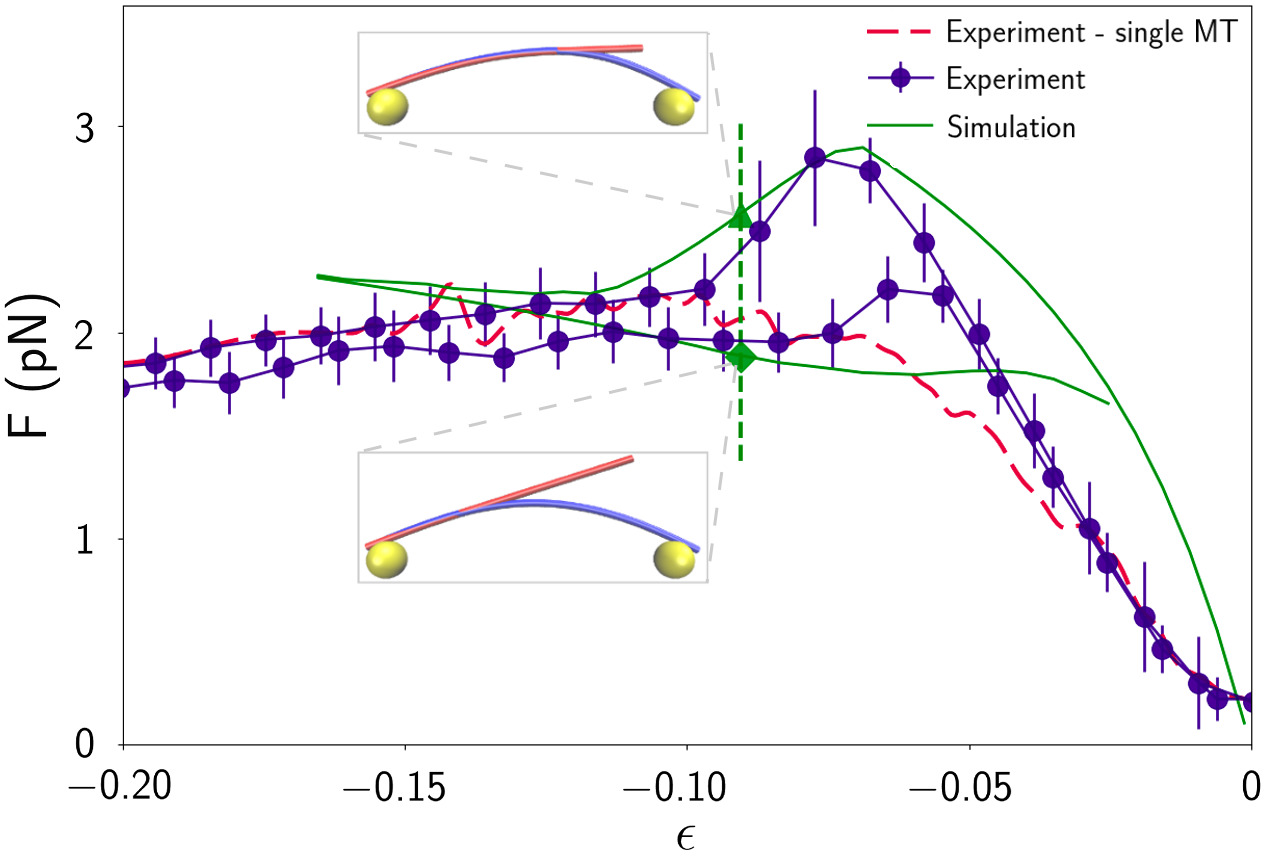}
\caption{ Measured force versus strain (blue) for a MT bundle composed of two MTs ($8.2 \,\mu \mathrm{m}$ and $5.6\, \mu \mathrm{m}$). The buckling curve of the longer microtubule alone  is shown in red. Simulation results are shown in green ($B = 19 \, \mathrm{pN} \, \mu \mathrm{m}^2$, $n = 100$, $K = 40 \, \mathrm{pN} / \mu \mathrm{m}  $, and $y_c = 0.02 \, \mu \mathrm{m}$) with filament configurations shown at two points of equal strain (green symbols).}
\end{figure}

\section{Hysteresis in buckled microtubule bundles: peeling from a curved substrate}

\subsection{Microtubule experiments} 

To further test our theory in a completely different setting, we chose to consider the adhesion between stiff cytoskeletal polymers, microtubules (MTs). We designed and conducted experiments involving a pair of microtubules (MTs) held together by the depletion interaction, induced by addition of non-adsorbing polymers. The range and strength of the tunable depletion attraction between the filaments is determined respectively by the size and the concentration of the polymer \citep{asakura1954, hilitski2015}. 

To obtain our bundled MT system, we start by using optical tweezers to attach micron-sized silica beads at two points along a single MT as described elsewhere \citep{memet2018}. Next, we attach a shorter MT to the longer filament by the depletion interaction and we link it to one bead by the biotin-streptavidin linkage (Fig. 2A). The mobile optical traps are displaced quasi-statically, subjecting the composite bundle to buckling forces that are measured using conventional techniques \citep{memet2018}. While the adhering MTs initially buckle together, above a critical strain the free end of the shorter MT begins to detach (``fray''). Further increasing strain leads to almost complete peeling of the shorter MT (Fig. 2A). From this point on, only the longer MT contributes to the buckling force, which is roughly independent of strain due to the effective softening induced by cross-sectional flattening \citep{memet2018,cross2019}. Reversing optical trap displacement reduces strain, eventually leading to re-adhesion, albeit at smaller curvatures/strains than for peeling. Hysteresis is apparent in the force-strain curves associated with this measurement, where strain $\epsilon = (d - L)/L$ (Fig. 2B).

Compared to peeling from a flat substrate, the microtubule system exhibits added complexity, as both microtubules are allowed to bend. Consequently, we need to revisit the theory, generalizing it for flexible substrates. Towards that end, we examine an elastic chain model in which both filaments are mobile and we can apply a buckling force at one end, through a bead attached to the longer filament (Fig. 3A). Letting functions $\kappa \left(s\right)$ and $\kappa' \left(s\right)$ characterize the curvatures of the two filaments, our previous results still hold (e.g. Eq. 4), but for relative curvature $\kappa_r \left(s\right) = | \kappa\left(s\right) - \kappa'\left(s\right)|$, whose maximal value $\kappa_c$ determines the onset of fraying. Previously, $\kappa' = 0$, $\kappa_r = \kappa$ and the bending energy scale could be expressed in terms of $\kappa_c$: $E_b^0 \sim B \kappa_c^2 l_H$. Here, the proper bending energy scale $E_b^0$ is not related to the relative curvature, but can be instead expressed as  $E_b^0 = B/L$ (force $B/L^2$ times length $L$). Meanwhile, we denote by $\widetilde E_b^0$ the energy scale $B \kappa_c^2 l_H$ and deem it the ``relative'' bending energy, since $\kappa_c$ now refers to the relative curvature $\kappa_r$. 

\subsection{Quantification of hysteresis and parameter scalings}

If we assume, in analogy with results from our first model, that the net energy loss $\delta E  \sim B \kappa_c^2 l_H \times \left(\Delta l/l_H\right)^2$  and express the dimensionless energy loss $n \, \delta e = n \, \delta E / E_b^0$ in terms of adhesion energy per unit length $J$, we get

\begin{equation}
n \, \delta e \sim \frac{L^3 \kappa_c^{5/2}}{n \, y_c^{1/2}},
\end{equation}

\noindent where $\kappa_c$ and $L$ are invariants. Therefore, invariance of $n \, \delta e$ requires that  $y_c \sim 1/n^2 $. Furthermore, we also need $K \sim n^3$ to keep $J$ invariant. Simulations (using the same molecular dynamics setup as in the 1D model in \cite{memet2018} and adding a second filament and a breakable adhesive interaction between corresponding beads on each filament -- see SM for details) confirm that the scaling forms $K \rightarrow n^3 K$ and $y_c \rightarrow y_c/n^2$ preserve hysteresis\footnote{provided we adjust the length of the shorter filament to account for the changing healing length $l_H \sim 1/n$, since the filaments are weakly bonded over this length scale.} (Fig. 3B, 3C). Moreover, plugging in the previously measured value $J \sim 0.1 $ pN for MT depletion-induced cohesion \citep{hilitski2015} allows us to reproduce both the onset of fraying and approximate hysteresis size (Fig. 4, green). Notably, we ignored factors such as the cross-sectional flattening of MTs~\cite{memet2018} and the hysteresis-narrowing effect of thermal fluctuations (Fig. S7), which would likely further improve the fit.

\section{Comparison between theory and experiments}

Equations (9) and (10) express hysteresis for the flat and curved substrate cases in terms of system parameters such as length $L$, interaction range $y_c$, adhesion strength $J$, and flexural rigidity $B$ (the latter two entering through $\kappa_c = \sqrt{2J/B}$). This allows us to predict, control, or compare the hysteresis of different systems. For instance, we notice that there is significantly less hysteresis in our microtubule experiments than in the graphene peeling experiments~\cite{miskin2017}. This observation can be understood in the context of our model by examining equations (9) and (10) for hysteresis in the graphene sheet (flat) and microtubule bundle (curved), respectively. Hysteresis in both cases is proportional to the adhesive length $L$ and with $\kappa_c = \sqrt{2J/B}$ but inversely proportional to $y_c$. While the factor of $\sqrt{J/B}$ is comparable between the two experiments (see SM), the other two length scales are not. The length of the graphene sheet is slightly larger ($L^{\mathrm{gr}} = 60 \mu \mathrm{m} > L^{\mathrm{MT}} \sim 10 \mu \mathrm{m}$) and, most importantly, the interaction in the graphene experiments is much shorter-range: $y_c^{\mathrm{gr}} \sim  0.1 \,\, \mathrm{nm} \ll y_c^{\mathrm{MT}} \sim  15 \,\,  \mathrm{nm}$. Thus, the much larger hysteresis observed in graphene peeling experiments is primarily due to the much shorter interaction range in graphene versus microtubules. 

Physically, we can understand the inverse dependence of hysteresis on $y_c$ by noting that, everything else being constant (i.e. $L$, $B$, and $J$), reducing $y_c$ results in stiffer springs $K$ (since $J = n K y_c^2 / L$ is constant). When one of these springs break, it will have a relatively large effect on the stress redistribution which will cause the newly broken bond to ``jump'' significantly (as in the inset of the schematic in Fig. 1A). We can see how such a jump relates to hysteresis by noting that if the direction of the free end displacement is reversed, the newly broken bond will have to travel extra distance (due to the jump) in order to re-enter the interaction range.  In contrast, larger $y_c$ means weaker springs, less effect of bond breakage, and thus less of a jump in the position of the broken bond (and consequently, smaller hysteresis).

\section{Summary and outlook}

Our theory of rate-independent adhesion hysteresis provides a quantitative mechanism for previously unexplained results showcasing substantial static hysteresis in graphene peeling \citep{miskin2017} as well as in microtubule bundles. More generally, our
results are applicable to any adhesive elastic system driven either quasi-statically and thus are relevant to diverse fields including nanoscience (graphene), cellular biophysics (microtubules), active matter (bundle disintegration), or material science (lattice trapping). We have shown that adhesion hysteresis arises due to energy lost at transitions between metastable states \citep{maddalena2009,puglisi2013} and quantified the manner in which hysteresis depends on elastic and adhesion parameters of the system, both for the case of a simple geometry in which the substrate is fixed and for that of a complex geometry in which the substrate is allowed to deform.

Our manuscript also describes a new experimental approach to the measurement of adhesion hysteresis in bundled filamentous polymers. Since filamentous bundles are an essential structural motif of the cellular cytoskeleton as well as the basic building block of biosynthetic active matter, understanding how they fray and disintegrate is relevant for the rich non-equilibrium dynamics of one of the best understood experimental systems in active matter, that of microtubules interacting with motors.

Complementing the classic work of Thomson on lattice trapping \cite{thomson1971}, we also show how the size and strength of the adhesive springs must be scaled with $n$ in order to preserve hysteresis in the continuum limit $n \rightarrow \infty$, thus eliminating a long-standing paradox by emphasizing the nature of distinguished limits required. The practical significance of our hysteresis-preserving scaling with $n$ is that it enables the self-consistent simulation of hysteretic systems at different levels of granularity. For example, in the case of a system in which the adhesion interaction is characterized by some finite spacing $\Delta l$ as well as known stiffness $K$ and cutoff $y_c$, the length scale may be too small to simulate the system efficiently. In that case, we may speed up the computational model by coarse-graining, starting from the real adhesive parameters and scaling them according to the hysteresis-preserving scaling laws.

Interestingly, the hysteresis mechanism we propose bears some resemblance to a phenomenon in polymer fracture known as the Lake-Thomas effect \citep{lake1967}, which remains an active area of research \citep{wang2019} despite being proposed over 50 years ago. In the Lake-Thomas effect the energy required to rupture an elastomer is much larger than the energy to break the chains crossing the fracture plane \citep{hui2003} due to the energy loss when the stretched chains away from the fracture zone relax as the crack propagates \citep{zhao2014,chen1991,creton2016,hui2003,andrews1987,brown2007}. To some extent, this notion of stretching (quasi-)globally while breaking locally also features in our theory, wherein at any given moment adhesive bonds within a region of size $l_H$ are stretched whereas upon peeling or healing, energy is released from a single bond, spanning a region of size $\Delta l$.  Although we have focused on the case of normal loading, in the context of shear loading of soft adhesive bonds, such as might be relevant in sliding friction between dissimilar materials, our proposed mechanism might also serve to explain energy loss even when the contact zone moves quasi-statically \citep{persson2013}. All together, our results should prove important in facilitating the modeling and simulation of adhesive hysteresis in many quasi-continuum elastic systems in both passive and active settings.

\bibliography{fray}

\begin{thebibliography}{54}%
\makeatletter
\providecommand \@ifxundefined [1]{%
 \@ifx{#1\undefined}
}%
\providecommand \@ifnum [1]{%
 \ifnum #1\expandafter \@firstoftwo
 \else \expandafter \@secondoftwo
 \fi
}%
\providecommand \@ifx [1]{%
 \ifx #1\expandafter \@firstoftwo
 \else \expandafter \@secondoftwo
 \fi
}%
\providecommand \natexlab [1]{#1}%
\providecommand \enquote  [1]{``#1''}%
\providecommand \bibnamefont  [1]{#1}%
\providecommand \bibfnamefont [1]{#1}%
\providecommand \citenamefont [1]{#1}%
\providecommand \href@noop [0]{\@secondoftwo}%
\providecommand \href [0]{\begingroup \@sanitize@url \@href}%
\providecommand \@href[1]{\@@startlink{#1}\@@href}%
\providecommand \@@href[1]{\endgroup#1\@@endlink}%
\providecommand \@sanitize@url [0]{\catcode `\\12\catcode `\$12\catcode
  `\&12\catcode `\#12\catcode `\^12\catcode `\_12\catcode `\%12\relax}%
\providecommand \@@startlink[1]{}%
\providecommand \@@endlink[0]{}%
\providecommand \url  [0]{\begingroup\@sanitize@url \@url }%
\providecommand \@url [1]{\endgroup\@href {#1}{\urlprefix }}%
\providecommand \urlprefix  [0]{URL }%
\providecommand \Eprint [0]{\href }%
\providecommand \doibase [0]{http://dx.doi.org/}%
\providecommand \selectlanguage [0]{\@gobble}%
\providecommand \bibinfo  [0]{\@secondoftwo}%
\providecommand \bibfield  [0]{\@secondoftwo}%
\providecommand \translation [1]{[#1]}%
\providecommand \BibitemOpen [0]{}%
\providecommand \bibitemStop [0]{}%
\providecommand \bibitemNoStop [0]{.\EOS\space}%
\providecommand \EOS [0]{\spacefactor3000\relax}%
\providecommand \BibitemShut  [1]{\csname bibitem#1\endcsname}%
\let\auto@bib@innerbib\@empty
\bibitem [{\citenamefont {Miskin}\ \emph {et~al.}(2017)\citenamefont {Miskin},
  \citenamefont {Sun}, \citenamefont {Cohen}, \citenamefont {Dichtel},\ and\
  \citenamefont {McEuen}}]{miskin2017}%
  \BibitemOpen
  \bibfield  {author} {\bibinfo {author} {\bibfnamefont {M.~Z.}\ \bibnamefont
  {Miskin}}, \bibinfo {author} {\bibfnamefont {C.}~\bibnamefont {Sun}},
  \bibinfo {author} {\bibfnamefont {I.}~\bibnamefont {Cohen}}, \bibinfo
  {author} {\bibfnamefont {W.~R.}\ \bibnamefont {Dichtel}}, \ and\ \bibinfo
  {author} {\bibfnamefont {P.~L.}\ \bibnamefont {McEuen}},\ }\href@noop {}
  {\bibfield  {journal} {\bibinfo  {journal} {Nano letters}\ }\textbf {\bibinfo
  {volume} {18}},\ \bibinfo {pages} {449} (\bibinfo {year} {2017})}\BibitemShut
  {NoStop}%
\bibitem [{\citenamefont {Sasaki}\ \emph {et~al.}(2008)\citenamefont {Sasaki},
  \citenamefont {Toyoda}, \citenamefont {Itamura},\ and\ \citenamefont
  {Miura}}]{sasaki2008}%
  \BibitemOpen
  \bibfield  {author} {\bibinfo {author} {\bibfnamefont {N.}~\bibnamefont
  {Sasaki}}, \bibinfo {author} {\bibfnamefont {A.}~\bibnamefont {Toyoda}},
  \bibinfo {author} {\bibfnamefont {N.}~\bibnamefont {Itamura}}, \ and\
  \bibinfo {author} {\bibfnamefont {K.}~\bibnamefont {Miura}},\ }\href@noop {}
  {\bibfield  {journal} {\bibinfo  {journal} {e-J Surf Sci Nanotechnol}\
  }\textbf {\bibinfo {volume} {6}},\ \bibinfo {pages} {72} (\bibinfo {year}
  {2008})}\BibitemShut {NoStop}%
\bibitem [{\citenamefont {Hosek}\ and\ \citenamefont {Tang}(2004)}]{hosek2004}%
  \BibitemOpen
  \bibfield  {author} {\bibinfo {author} {\bibfnamefont {M.}~\bibnamefont
  {Hosek}}\ and\ \bibinfo {author} {\bibfnamefont {J.}~\bibnamefont {Tang}},\
  }\href@noop {} {\bibfield  {journal} {\bibinfo  {journal} {Phys. Rev. E}\
  }\textbf {\bibinfo {volume} {69}},\ \bibinfo {pages} {051907} (\bibinfo
  {year} {2004})}\BibitemShut {NoStop}%
\bibitem [{\citenamefont {Whitelam}\ \emph {et~al.}(2008)\citenamefont
  {Whitelam}, \citenamefont {Pronk},\ and\ \citenamefont
  {Geissler}}]{whitelam2008}%
  \BibitemOpen
  \bibfield  {author} {\bibinfo {author} {\bibfnamefont {S.}~\bibnamefont
  {Whitelam}}, \bibinfo {author} {\bibfnamefont {S.}~\bibnamefont {Pronk}}, \
  and\ \bibinfo {author} {\bibfnamefont {P.~L.}\ \bibnamefont {Geissler}},\
  }\href@noop {} {\bibfield  {journal} {\bibinfo  {journal} {Biophysical
  Journal}\ }\textbf {\bibinfo {volume} {94}},\ \bibinfo {pages} {2452}
  (\bibinfo {year} {2008})}\BibitemShut {NoStop}%
\bibitem [{\citenamefont {Smith}\ \emph {et~al.}(1996)\citenamefont {Smith},
  \citenamefont {Cui},\ and\ \citenamefont {Bustamante}}]{smith1996}%
  \BibitemOpen
  \bibfield  {author} {\bibinfo {author} {\bibfnamefont {S.~B.}\ \bibnamefont
  {Smith}}, \bibinfo {author} {\bibfnamefont {Y.}~\bibnamefont {Cui}}, \ and\
  \bibinfo {author} {\bibfnamefont {C.}~\bibnamefont {Bustamante}},\
  }\href@noop {} {\bibfield  {journal} {\bibinfo  {journal} {Science}\ }\textbf
  {\bibinfo {volume} {271}},\ \bibinfo {pages} {795} (\bibinfo {year}
  {1996})}\BibitemShut {NoStop}%
\bibitem [{\citenamefont {Rouzina}\ and\ \citenamefont
  {Bloomfield}(2001)}]{rouzina2001}%
  \BibitemOpen
  \bibfield  {author} {\bibinfo {author} {\bibfnamefont {I.}~\bibnamefont
  {Rouzina}}\ and\ \bibinfo {author} {\bibfnamefont {V.~A.}\ \bibnamefont
  {Bloomfield}},\ }\href@noop {} {\bibfield  {journal} {\bibinfo  {journal}
  {Biophysical journal}\ }\textbf {\bibinfo {volume} {80}},\ \bibinfo {pages}
  {882} (\bibinfo {year} {2001})}\BibitemShut {NoStop}%
\bibitem [{\citenamefont {Peyrard}(2004)}]{peyrard2004}%
  \BibitemOpen
  \bibfield  {author} {\bibinfo {author} {\bibfnamefont {M.}~\bibnamefont
  {Peyrard}},\ }\href@noop {} {\bibfield  {journal} {\bibinfo  {journal}
  {Nonlinearity}\ }\textbf {\bibinfo {volume} {17}},\ \bibinfo {pages} {R1}
  (\bibinfo {year} {2004})}\BibitemShut {NoStop}%
\bibitem [{\citenamefont {Dembo}\ \emph {et~al.}(1988)\citenamefont {Dembo},
  \citenamefont {Torney}, \citenamefont {Saxman},\ and\ \citenamefont
  {Hammer}}]{dembo1988}%
  \BibitemOpen
  \bibfield  {author} {\bibinfo {author} {\bibfnamefont {M.}~\bibnamefont
  {Dembo}}, \bibinfo {author} {\bibfnamefont {D.}~\bibnamefont {Torney}},
  \bibinfo {author} {\bibfnamefont {K.}~\bibnamefont {Saxman}}, \ and\ \bibinfo
  {author} {\bibfnamefont {D.}~\bibnamefont {Hammer}},\ }\href@noop {}
  {\bibfield  {journal} {\bibinfo  {journal} {Proc. R. Soc. Lond. B}\ }\textbf
  {\bibinfo {volume} {234}},\ \bibinfo {pages} {55} (\bibinfo {year}
  {1988})}\BibitemShut {NoStop}%
\bibitem [{\citenamefont {Aoyama}\ and\ \citenamefont
  {Kamiya}(2005)}]{aoyama2005}%
  \BibitemOpen
  \bibfield  {author} {\bibinfo {author} {\bibfnamefont {S.}~\bibnamefont
  {Aoyama}}\ and\ \bibinfo {author} {\bibfnamefont {R.}~\bibnamefont
  {Kamiya}},\ }\href@noop {} {\bibfield  {journal} {\bibinfo  {journal}
  {Biophys. J.}\ }\textbf {\bibinfo {volume} {89}},\ \bibinfo {pages} {3261}
  (\bibinfo {year} {2005})}\BibitemShut {NoStop}%
\bibitem [{\citenamefont {Sanchez}\ \emph {et~al.}(2012)\citenamefont
  {Sanchez}, \citenamefont {Chen}, \citenamefont {DeCamp}, \citenamefont
  {Heymann},\ and\ \citenamefont {Dogic}}]{sanchez2012spontaneous}%
  \BibitemOpen
  \bibfield  {author} {\bibinfo {author} {\bibfnamefont {T.}~\bibnamefont
  {Sanchez}}, \bibinfo {author} {\bibfnamefont {D.~T.}\ \bibnamefont {Chen}},
  \bibinfo {author} {\bibfnamefont {S.~J.}\ \bibnamefont {DeCamp}}, \bibinfo
  {author} {\bibfnamefont {M.}~\bibnamefont {Heymann}}, \ and\ \bibinfo
  {author} {\bibfnamefont {Z.}~\bibnamefont {Dogic}},\ }\href@noop {}
  {\bibfield  {journal} {\bibinfo  {journal} {Nature}\ }\textbf {\bibinfo
  {volume} {491}},\ \bibinfo {pages} {431} (\bibinfo {year}
  {2012})}\BibitemShut {NoStop}%
\bibitem [{\citenamefont {Maugis}(2013)}]{maugis2013}%
  \BibitemOpen
  \bibfield  {author} {\bibinfo {author} {\bibfnamefont {D.}~\bibnamefont
  {Maugis}},\ }\href@noop {} {\emph {\bibinfo {title} {Contact, adhesion and
  rupture of elastic solids}}},\ Vol.\ \bibinfo {volume} {130}\ (\bibinfo
  {publisher} {Springer},\ \bibinfo {year} {2013})\BibitemShut {NoStop}%
\bibitem [{\citenamefont {Pesika}\ \emph {et~al.}(2007)\citenamefont {Pesika},
  \citenamefont {Tian}, \citenamefont {Zhao}, \citenamefont {Rosenberg},
  \citenamefont {Zeng}, \citenamefont {McGuiggan}, \citenamefont {Autumn},\
  and\ \citenamefont {Israelachvili}}]{pesika2007}%
  \BibitemOpen
  \bibfield  {author} {\bibinfo {author} {\bibfnamefont {N.~S.}\ \bibnamefont
  {Pesika}}, \bibinfo {author} {\bibfnamefont {Y.}~\bibnamefont {Tian}},
  \bibinfo {author} {\bibfnamefont {B.}~\bibnamefont {Zhao}}, \bibinfo {author}
  {\bibfnamefont {K.}~\bibnamefont {Rosenberg}}, \bibinfo {author}
  {\bibfnamefont {H.}~\bibnamefont {Zeng}}, \bibinfo {author} {\bibfnamefont
  {P.}~\bibnamefont {McGuiggan}}, \bibinfo {author} {\bibfnamefont
  {K.}~\bibnamefont {Autumn}}, \ and\ \bibinfo {author} {\bibfnamefont {J.~N.}\
  \bibnamefont {Israelachvili}},\ }\href@noop {} {\bibfield  {journal}
  {\bibinfo  {journal} {J. Adhes.}\ }\textbf {\bibinfo {volume} {83}},\
  \bibinfo {pages} {383} (\bibinfo {year} {2007})}\BibitemShut {NoStop}%
\bibitem [{\citenamefont {Molinari}\ and\ \citenamefont
  {Ravichandran}(2008)}]{molinari2008}%
  \BibitemOpen
  \bibfield  {author} {\bibinfo {author} {\bibfnamefont {A.}~\bibnamefont
  {Molinari}}\ and\ \bibinfo {author} {\bibfnamefont {G.}~\bibnamefont
  {Ravichandran}},\ }\href@noop {} {\bibfield  {journal} {\bibinfo  {journal}
  {J. Adhes.}\ }\textbf {\bibinfo {volume} {84}},\ \bibinfo {pages} {961}
  (\bibinfo {year} {2008})}\BibitemShut {NoStop}%
\bibitem [{\citenamefont {Cohen}\ \emph {et~al.}(2018)\citenamefont {Cohen},
  \citenamefont {Chan},\ and\ \citenamefont {Mahadevan}}]{cohen2018}%
  \BibitemOpen
  \bibfield  {author} {\bibinfo {author} {\bibfnamefont {T.}~\bibnamefont
  {Cohen}}, \bibinfo {author} {\bibfnamefont {C.~U.}\ \bibnamefont {Chan}}, \
  and\ \bibinfo {author} {\bibfnamefont {L.}~\bibnamefont {Mahadevan}},\
  }\href@noop {} {\bibfield  {journal} {\bibinfo  {journal} {Soft matter}\
  }\textbf {\bibinfo {volume} {14}},\ \bibinfo {pages} {1771} (\bibinfo {year}
  {2018})}\BibitemShut {NoStop}%
\bibitem [{\citenamefont {Williams}(2014)}]{williams2014}%
  \BibitemOpen
  \bibfield  {author} {\bibinfo {author} {\bibfnamefont {J.~A.}\ \bibnamefont
  {Williams}},\ }\href@noop {} {\bibfield  {journal} {\bibinfo  {journal} {J.
  Phys. D: Appl. Phys.}\ }\textbf {\bibinfo {volume} {48}},\ \bibinfo {pages}
  {015401} (\bibinfo {year} {2014})}\BibitemShut {NoStop}%
\bibitem [{\citenamefont {Obreimoff}(1930)}]{obreimoff1930}%
  \BibitemOpen
  \bibfield  {author} {\bibinfo {author} {\bibfnamefont {J.}~\bibnamefont
  {Obreimoff}},\ }\href@noop {} {\bibfield  {journal} {\bibinfo  {journal}
  {Proc. R. Soc. Lond. A}\ }\textbf {\bibinfo {volume} {127}},\ \bibinfo
  {pages} {290} (\bibinfo {year} {1930})}\BibitemShut {NoStop}%
\bibitem [{\citenamefont {Kendall}(2007)}]{kendall2007}%
  \BibitemOpen
  \bibfield  {author} {\bibinfo {author} {\bibfnamefont {K.}~\bibnamefont
  {Kendall}},\ }\href@noop {} {\emph {\bibinfo {title} {Molecular adhesion and
  its applications: the sticky universe}}}\ (\bibinfo  {publisher} {Springer},\
  \bibinfo {year} {2007})\BibitemShut {NoStop}%
\bibitem [{\citenamefont {Pocius}\ and\ \citenamefont
  {Dillard}(2002)}]{pocius2002}%
  \BibitemOpen
  \bibfield  {author} {\bibinfo {author} {\bibfnamefont {A.~V.}\ \bibnamefont
  {Pocius}}\ and\ \bibinfo {author} {\bibfnamefont {D.~A.}\ \bibnamefont
  {Dillard}},\ }\href@noop {} {\emph {\bibinfo {title} {Adhesion science and
  engineering: surfaces, chemistry and applications}}}\ (\bibinfo  {publisher}
  {Elsevier},\ \bibinfo {year} {2002})\BibitemShut {NoStop}%
\bibitem [{\citenamefont {Liu}\ \emph {et~al.}(2018)\citenamefont {Liu},
  \citenamefont {Lu}, \citenamefont {Zheng}, \citenamefont {Tao}, \citenamefont
  {Meng},\ and\ \citenamefont {Tian}}]{liu2018}%
  \BibitemOpen
  \bibfield  {author} {\bibinfo {author} {\bibfnamefont {Z.}~\bibnamefont
  {Liu}}, \bibinfo {author} {\bibfnamefont {H.}~\bibnamefont {Lu}}, \bibinfo
  {author} {\bibfnamefont {Y.}~\bibnamefont {Zheng}}, \bibinfo {author}
  {\bibfnamefont {D.}~\bibnamefont {Tao}}, \bibinfo {author} {\bibfnamefont
  {Y.}~\bibnamefont {Meng}}, \ and\ \bibinfo {author} {\bibfnamefont
  {Y.}~\bibnamefont {Tian}},\ }\href@noop {} {\bibfield  {journal} {\bibinfo
  {journal} {Sci. reports}\ }\textbf {\bibinfo {volume} {8}},\ \bibinfo {pages}
  {6147} (\bibinfo {year} {2018})}\BibitemShut {NoStop}%
\bibitem [{\citenamefont {Chung}\ and\ \citenamefont
  {Chaudhury}(2005)}]{chung2005}%
  \BibitemOpen
  \bibfield  {author} {\bibinfo {author} {\bibfnamefont {J.~Y.}\ \bibnamefont
  {Chung}}\ and\ \bibinfo {author} {\bibfnamefont {M.~K.}\ \bibnamefont
  {Chaudhury}},\ }\href@noop {} {\bibfield  {journal} {\bibinfo  {journal} {J.
  R. Soc. Interface}\ }\textbf {\bibinfo {volume} {2}},\ \bibinfo {pages} {55}
  (\bibinfo {year} {2005})}\BibitemShut {NoStop}%
\bibitem [{\citenamefont {De~Zotti}\ \emph {et~al.}(2019)\citenamefont
  {De~Zotti}, \citenamefont {Rapina}, \citenamefont {Cortet}, \citenamefont
  {Vanel},\ and\ \citenamefont {Santucci}}]{de2019}%
  \BibitemOpen
  \bibfield  {author} {\bibinfo {author} {\bibfnamefont {V.}~\bibnamefont
  {De~Zotti}}, \bibinfo {author} {\bibfnamefont {K.}~\bibnamefont {Rapina}},
  \bibinfo {author} {\bibfnamefont {P.-P.}\ \bibnamefont {Cortet}}, \bibinfo
  {author} {\bibfnamefont {L.}~\bibnamefont {Vanel}}, \ and\ \bibinfo {author}
  {\bibfnamefont {S.}~\bibnamefont {Santucci}},\ }\href@noop {} {\bibfield
  {journal} {\bibinfo  {journal} {Phys. Rev. Lett.}\ }\textbf {\bibinfo
  {volume} {122}},\ \bibinfo {pages} {068005} (\bibinfo {year}
  {2019})}\BibitemShut {NoStop}%
\bibitem [{\citenamefont {Villey}\ \emph {et~al.}(2015)\citenamefont {Villey},
  \citenamefont {Creton}, \citenamefont {Cortet}, \citenamefont {Dalbe},
  \citenamefont {Jet}, \citenamefont {Saintyves}, \citenamefont {Santucci},
  \citenamefont {Vanel}, \citenamefont {Yarusso},\ and\ \citenamefont
  {Ciccotti}}]{villey2015}%
  \BibitemOpen
  \bibfield  {author} {\bibinfo {author} {\bibfnamefont {R.}~\bibnamefont
  {Villey}}, \bibinfo {author} {\bibfnamefont {C.}~\bibnamefont {Creton}},
  \bibinfo {author} {\bibfnamefont {P.-P.}\ \bibnamefont {Cortet}}, \bibinfo
  {author} {\bibfnamefont {M.-J.}\ \bibnamefont {Dalbe}}, \bibinfo {author}
  {\bibfnamefont {T.}~\bibnamefont {Jet}}, \bibinfo {author} {\bibfnamefont
  {B.}~\bibnamefont {Saintyves}}, \bibinfo {author} {\bibfnamefont
  {S.}~\bibnamefont {Santucci}}, \bibinfo {author} {\bibfnamefont
  {L.}~\bibnamefont {Vanel}}, \bibinfo {author} {\bibfnamefont {D.~J.}\
  \bibnamefont {Yarusso}}, \ and\ \bibinfo {author} {\bibfnamefont
  {M.}~\bibnamefont {Ciccotti}},\ }\href@noop {} {\bibfield  {journal}
  {\bibinfo  {journal} {Soft Matter}\ }\textbf {\bibinfo {volume} {11}},\
  \bibinfo {pages} {3480} (\bibinfo {year} {2015})}\BibitemShut {NoStop}%
\bibitem [{\citenamefont {Chen}\ \emph {et~al.}(1991)\citenamefont {Chen},
  \citenamefont {Helm},\ and\ \citenamefont {Israelachvili}}]{chen1991}%
  \BibitemOpen
  \bibfield  {author} {\bibinfo {author} {\bibfnamefont {Y.}~\bibnamefont
  {Chen}}, \bibinfo {author} {\bibfnamefont {C.}~\bibnamefont {Helm}}, \ and\
  \bibinfo {author} {\bibfnamefont {J.}~\bibnamefont {Israelachvili}},\
  }\href@noop {} {\bibfield  {journal} {\bibinfo  {journal} {J. Phys. Chem.}\
  }\textbf {\bibinfo {volume} {95}},\ \bibinfo {pages} {10736} (\bibinfo {year}
  {1991})}\BibitemShut {NoStop}%
\bibitem [{\citenamefont {Sekiguchi}\ \emph {et~al.}(2014)\citenamefont
  {Sekiguchi}, \citenamefont {Hemthavy}, \citenamefont {Saito},\ and\
  \citenamefont {Takahashi}}]{sekiguchi2014}%
  \BibitemOpen
  \bibfield  {author} {\bibinfo {author} {\bibfnamefont {Y.}~\bibnamefont
  {Sekiguchi}}, \bibinfo {author} {\bibfnamefont {P.}~\bibnamefont {Hemthavy}},
  \bibinfo {author} {\bibfnamefont {S.}~\bibnamefont {Saito}}, \ and\ \bibinfo
  {author} {\bibfnamefont {K.}~\bibnamefont {Takahashi}},\ }\href@noop {}
  {\bibfield  {journal} {\bibinfo  {journal} {Int. J. Adhes. Adhes.}\ }\textbf
  {\bibinfo {volume} {49}},\ \bibinfo {pages} {1} (\bibinfo {year}
  {2014})}\BibitemShut {NoStop}%
\bibitem [{\citenamefont {Puglisi}\ and\ \citenamefont
  {Truskinovsky}(2013)}]{puglisi2013}%
  \BibitemOpen
  \bibfield  {author} {\bibinfo {author} {\bibfnamefont {G.}~\bibnamefont
  {Puglisi}}\ and\ \bibinfo {author} {\bibfnamefont {L.}~\bibnamefont
  {Truskinovsky}},\ }\href@noop {} {\bibfield  {journal} {\bibinfo  {journal}
  {Phys. Rev. E}\ }\textbf {\bibinfo {volume} {87}},\ \bibinfo {pages} {032714}
  (\bibinfo {year} {2013})}\BibitemShut {NoStop}%
\bibitem [{\citenamefont {Maddalena}\ \emph {et~al.}(2009)\citenamefont
  {Maddalena}, \citenamefont {Percivale}, \citenamefont {Puglisi},\ and\
  \citenamefont {Truskinovsky}}]{maddalena2009}%
  \BibitemOpen
  \bibfield  {author} {\bibinfo {author} {\bibfnamefont {F.}~\bibnamefont
  {Maddalena}}, \bibinfo {author} {\bibfnamefont {D.}~\bibnamefont
  {Percivale}}, \bibinfo {author} {\bibfnamefont {G.}~\bibnamefont {Puglisi}},
  \ and\ \bibinfo {author} {\bibfnamefont {L.}~\bibnamefont {Truskinovsky}},\
  }\href@noop {} {\bibfield  {journal} {\bibinfo  {journal} {Cont. Mech.
  Therm.}\ }\textbf {\bibinfo {volume} {21}},\ \bibinfo {pages} {251} (\bibinfo
  {year} {2009})}\BibitemShut {NoStop}%
\bibitem [{\citenamefont {Evans}(1985{\natexlab{a}})}]{evans1985a}%
  \BibitemOpen
  \bibfield  {author} {\bibinfo {author} {\bibfnamefont {E.~A.}\ \bibnamefont
  {Evans}},\ }\href@noop {} {\bibfield  {journal} {\bibinfo  {journal}
  {Biophysical journal}\ }\textbf {\bibinfo {volume} {48}},\ \bibinfo {pages}
  {175} (\bibinfo {year} {1985}{\natexlab{a}})}\BibitemShut {NoStop}%
\bibitem [{\citenamefont {Evans}(1985{\natexlab{b}})}]{evans1985b}%
  \BibitemOpen
  \bibfield  {author} {\bibinfo {author} {\bibfnamefont {E.}~\bibnamefont
  {Evans}},\ }\href@noop {} {\bibfield  {journal} {\bibinfo  {journal}
  {Biophysical Journal}\ }\textbf {\bibinfo {volume} {48}},\ \bibinfo {pages}
  {185} (\bibinfo {year} {1985}{\natexlab{b}})}\BibitemShut {NoStop}%
\bibitem [{\citenamefont {Thomson}\ \emph {et~al.}(1971)\citenamefont
  {Thomson}, \citenamefont {Hsieh},\ and\ \citenamefont {Rana}}]{thomson1971}%
  \BibitemOpen
  \bibfield  {author} {\bibinfo {author} {\bibfnamefont {R.}~\bibnamefont
  {Thomson}}, \bibinfo {author} {\bibfnamefont {C.}~\bibnamefont {Hsieh}}, \
  and\ \bibinfo {author} {\bibfnamefont {V.}~\bibnamefont {Rana}},\ }\href@noop
  {} {\bibfield  {journal} {\bibinfo  {journal} {J. Appl. Phys.}\ }\textbf
  {\bibinfo {volume} {42}},\ \bibinfo {pages} {3154} (\bibinfo {year}
  {1971})}\BibitemShut {NoStop}%
\bibitem [{\citenamefont {Gao}\ and\ \citenamefont {Rice}(1989)}]{gao1989}%
  \BibitemOpen
  \bibfield  {author} {\bibinfo {author} {\bibfnamefont {H.}~\bibnamefont
  {Gao}}\ and\ \bibinfo {author} {\bibfnamefont {J.~R.}\ \bibnamefont {Rice}},\
  }\href@noop {} {\  (\bibinfo {year} {1989})}\BibitemShut {NoStop}%
\bibitem [{\citenamefont {Rice}(1978)}]{rice1978}%
  \BibitemOpen
  \bibfield  {author} {\bibinfo {author} {\bibfnamefont {J.}~\bibnamefont
  {Rice}},\ }\href@noop {} {\bibfield  {journal} {\bibinfo  {journal} {Journal
  of the Mechanics and Physics of Solids}\ }\textbf {\bibinfo {volume} {26}},\
  \bibinfo {pages} {61} (\bibinfo {year} {1978})}\BibitemShut {NoStop}%
\bibitem [{\citenamefont {Guduru}(2007)}]{guduru2007}%
  \BibitemOpen
  \bibfield  {author} {\bibinfo {author} {\bibfnamefont {P.}~\bibnamefont
  {Guduru}},\ }\href@noop {} {\bibfield  {journal} {\bibinfo  {journal}
  {Journal of the Mechanics and Physics of Solids}\ }\textbf {\bibinfo {volume}
  {55}},\ \bibinfo {pages} {445} (\bibinfo {year} {2007})}\BibitemShut
  {NoStop}%
\bibitem [{\citenamefont {Noderer}\ \emph {et~al.}(2007)\citenamefont
  {Noderer}, \citenamefont {Shen}, \citenamefont {Vajpayee}, \citenamefont
  {Glassmaker}, \citenamefont {Jagota},\ and\ \citenamefont
  {Hui}}]{noderer2007}%
  \BibitemOpen
  \bibfield  {author} {\bibinfo {author} {\bibfnamefont {W.}~\bibnamefont
  {Noderer}}, \bibinfo {author} {\bibfnamefont {L.}~\bibnamefont {Shen}},
  \bibinfo {author} {\bibfnamefont {S.}~\bibnamefont {Vajpayee}}, \bibinfo
  {author} {\bibfnamefont {N.}~\bibnamefont {Glassmaker}}, \bibinfo {author}
  {\bibfnamefont {A.}~\bibnamefont {Jagota}}, \ and\ \bibinfo {author}
  {\bibfnamefont {C.-Y.}\ \bibnamefont {Hui}},\ }\href@noop {} {\bibfield
  {journal} {\bibinfo  {journal} {Proceedings of the Royal Society A:
  Mathematical, Physical and Engineering Sciences}\ }\textbf {\bibinfo {volume}
  {463}},\ \bibinfo {pages} {2631} (\bibinfo {year} {2007})}\BibitemShut
  {NoStop}%
\bibitem [{\citenamefont {Kesari}\ and\ \citenamefont
  {Lew}(2011)}]{kesari2011}%
  \BibitemOpen
  \bibfield  {author} {\bibinfo {author} {\bibfnamefont {H.}~\bibnamefont
  {Kesari}}\ and\ \bibinfo {author} {\bibfnamefont {A.~J.}\ \bibnamefont
  {Lew}},\ }\href@noop {} {\bibfield  {journal} {\bibinfo  {journal} {Journal
  of the Mechanics and Physics of Solids}\ }\textbf {\bibinfo {volume} {59}},\
  \bibinfo {pages} {2488} (\bibinfo {year} {2011})}\BibitemShut {NoStop}%
\bibitem [{\citenamefont {Xia}\ \emph {et~al.}(2013)\citenamefont {Xia},
  \citenamefont {Ponson}, \citenamefont {Ravichandran},\ and\ \citenamefont
  {Bhattacharya}}]{xia2013}%
  \BibitemOpen
  \bibfield  {author} {\bibinfo {author} {\bibfnamefont {S.}~\bibnamefont
  {Xia}}, \bibinfo {author} {\bibfnamefont {L.}~\bibnamefont {Ponson}},
  \bibinfo {author} {\bibfnamefont {G.}~\bibnamefont {Ravichandran}}, \ and\
  \bibinfo {author} {\bibfnamefont {K.}~\bibnamefont {Bhattacharya}},\
  }\href@noop {} {\bibfield  {journal} {\bibinfo  {journal} {Journal of the
  Mechanics and Physics of Solids}\ }\textbf {\bibinfo {volume} {61}},\
  \bibinfo {pages} {838} (\bibinfo {year} {2013})}\BibitemShut {NoStop}%
\bibitem [{\citenamefont {Kendall}(1975)}]{kendall1975}%
  \BibitemOpen
  \bibfield  {author} {\bibinfo {author} {\bibfnamefont {K.}~\bibnamefont
  {Kendall}},\ }\href@noop {} {\bibfield  {journal} {\bibinfo  {journal}
  {Proceedings of the Royal Society of London. A. Mathematical and Physical
  Sciences}\ }\textbf {\bibinfo {volume} {341}},\ \bibinfo {pages} {409}
  (\bibinfo {year} {1975})}\BibitemShut {NoStop}%
\bibitem [{\citenamefont {Zapperi}\ and\ \citenamefont
  {Mahadevan}(2011)}]{zapperi2011}%
  \BibitemOpen
  \bibfield  {author} {\bibinfo {author} {\bibfnamefont {S.}~\bibnamefont
  {Zapperi}}\ and\ \bibinfo {author} {\bibfnamefont {L.}~\bibnamefont
  {Mahadevan}},\ }\href@noop {} {\bibfield  {journal} {\bibinfo  {journal}
  {Biophys. J.}\ }\textbf {\bibinfo {volume} {101}},\ \bibinfo {pages} {267}
  (\bibinfo {year} {2011})}\BibitemShut {NoStop}%
\bibitem [{\citenamefont {J{\'a}nosi}\ \emph {et~al.}(1998)\citenamefont
  {J{\'a}nosi}, \citenamefont {Chr{\'e}tien},\ and\ \citenamefont
  {Flyvbjerg}}]{janosi1998}%
  \BibitemOpen
  \bibfield  {author} {\bibinfo {author} {\bibfnamefont {I.~M.}\ \bibnamefont
  {J{\'a}nosi}}, \bibinfo {author} {\bibfnamefont {D.}~\bibnamefont
  {Chr{\'e}tien}}, \ and\ \bibinfo {author} {\bibfnamefont {H.}~\bibnamefont
  {Flyvbjerg}},\ }\href {\doibase 10.1007/s002490050160} {\bibfield  {journal}
  {\bibinfo  {journal} {Eur. Biophys. J.}\ }\textbf {\bibinfo {volume} {27}},\
  \bibinfo {pages} {501} (\bibinfo {year} {1998})}\BibitemShut {NoStop}%
\bibitem [{\citenamefont {Stewart}\ \emph {et~al.}(1987)\citenamefont
  {Stewart}, \citenamefont {McLachlan},\ and\ \citenamefont
  {Calladine}}]{stewart1987}%
  \BibitemOpen
  \bibfield  {author} {\bibinfo {author} {\bibfnamefont {M.}~\bibnamefont
  {Stewart}}, \bibinfo {author} {\bibfnamefont {A.~D.}\ \bibnamefont
  {McLachlan}}, \ and\ \bibinfo {author} {\bibfnamefont {C.~R.}\ \bibnamefont
  {Calladine}},\ }\href@noop {} {\bibfield  {journal} {\bibinfo  {journal}
  {Proc. R. Soc. Lond. B}\ }\textbf {\bibinfo {volume} {229}},\ \bibinfo
  {pages} {381} (\bibinfo {year} {1987})}\BibitemShut {NoStop}%
\bibitem [{\citenamefont {Majidi}\ \emph {et~al.}(2012)\citenamefont {Majidi},
  \citenamefont {O'Reilly},\ and\ \citenamefont {Williams}}]{majidi2012}%
  \BibitemOpen
  \bibfield  {author} {\bibinfo {author} {\bibfnamefont {C.}~\bibnamefont
  {Majidi}}, \bibinfo {author} {\bibfnamefont {O.~M.}\ \bibnamefont
  {O'Reilly}}, \ and\ \bibinfo {author} {\bibfnamefont {J.~A.}\ \bibnamefont
  {Williams}},\ }\href@noop {} {\bibfield  {journal} {\bibinfo  {journal} {J.
  Mech. Phys. Sol.}\ }\textbf {\bibinfo {volume} {60}},\ \bibinfo {pages} {827}
  (\bibinfo {year} {2012})}\BibitemShut {NoStop}%
\bibitem [{\citenamefont {Arnold}\ \emph {et~al.}(2013)\citenamefont {Arnold},
  \citenamefont {Lenz}, \citenamefont {Kesselheim}, \citenamefont {Weeber},
  \citenamefont {Fahrenberger}, \citenamefont {Roehm}, \citenamefont
  {Ko\v{s}ovan},\ and\ \citenamefont {Holm}}]{Arnold2013}%
  \BibitemOpen
  \bibfield  {author} {\bibinfo {author} {\bibfnamefont {A.}~\bibnamefont
  {Arnold}}, \bibinfo {author} {\bibfnamefont {O.}~\bibnamefont {Lenz}},
  \bibinfo {author} {\bibfnamefont {S.}~\bibnamefont {Kesselheim}}, \bibinfo
  {author} {\bibfnamefont {R.}~\bibnamefont {Weeber}}, \bibinfo {author}
  {\bibfnamefont {F.}~\bibnamefont {Fahrenberger}}, \bibinfo {author}
  {\bibfnamefont {D.}~\bibnamefont {Roehm}}, \bibinfo {author} {\bibfnamefont
  {P.}~\bibnamefont {Ko\v{s}ovan}}, \ and\ \bibinfo {author} {\bibfnamefont
  {C.}~\bibnamefont {Holm}},\ }in\ \href {\doibase 10.1007/978-3-642-32979-1_1}
  {\emph {\bibinfo {booktitle} {Meshfree Methods for Partial Differential
  Equations {VI}}}}\ (\bibinfo  {publisher} {Springer},\ \bibinfo {year}
  {2013})\ pp.\ \bibinfo {pages} {1--23}\BibitemShut {NoStop}%
\bibitem [{\citenamefont {Asakura}\ and\ \citenamefont
  {Oosawa}(1954)}]{asakura1954}%
  \BibitemOpen
  \bibfield  {author} {\bibinfo {author} {\bibfnamefont {S.}~\bibnamefont
  {Asakura}}\ and\ \bibinfo {author} {\bibfnamefont {F.}~\bibnamefont
  {Oosawa}},\ }\href@noop {} {\bibfield  {journal} {\bibinfo  {journal} {J.
  Chem. Phys.}\ }\textbf {\bibinfo {volume} {22}},\ \bibinfo {pages} {1255}
  (\bibinfo {year} {1954})}\BibitemShut {NoStop}%
\bibitem [{\citenamefont {Hilitski}\ \emph {et~al.}(2015)\citenamefont
  {Hilitski}, \citenamefont {Ward}, \citenamefont {Cajamarca}, \citenamefont
  {Hagan}, \citenamefont {Grason},\ and\ \citenamefont {Dogic}}]{hilitski2015}%
  \BibitemOpen
  \bibfield  {author} {\bibinfo {author} {\bibfnamefont {F.}~\bibnamefont
  {Hilitski}}, \bibinfo {author} {\bibfnamefont {A.~R.}\ \bibnamefont {Ward}},
  \bibinfo {author} {\bibfnamefont {L.}~\bibnamefont {Cajamarca}}, \bibinfo
  {author} {\bibfnamefont {M.~F.}\ \bibnamefont {Hagan}}, \bibinfo {author}
  {\bibfnamefont {G.~M.}\ \bibnamefont {Grason}}, \ and\ \bibinfo {author}
  {\bibfnamefont {Z.}~\bibnamefont {Dogic}},\ }\href@noop {} {\bibfield
  {journal} {\bibinfo  {journal} {Phys. Rev. Lett.}\ }\textbf {\bibinfo
  {volume} {114}},\ \bibinfo {pages} {138102} (\bibinfo {year}
  {2015})}\BibitemShut {NoStop}%
\bibitem [{\citenamefont {Memet}\ \emph {et~al.}(2018)\citenamefont {Memet},
  \citenamefont {Hilitsk}, \citenamefont {Morris}, \citenamefont {Schwenger},
  \citenamefont {Dogic},\ and\ \citenamefont {Mahadevan}}]{memet2018}%
  \BibitemOpen
  \bibfield  {author} {\bibinfo {author} {\bibfnamefont {E.}~\bibnamefont
  {Memet}}, \bibinfo {author} {\bibfnamefont {F.}~\bibnamefont {Hilitsk}},
  \bibinfo {author} {\bibfnamefont {M.~A.}\ \bibnamefont {Morris}}, \bibinfo
  {author} {\bibfnamefont {W.~J.}\ \bibnamefont {Schwenger}}, \bibinfo {author}
  {\bibfnamefont {Z.}~\bibnamefont {Dogic}}, \ and\ \bibinfo {author}
  {\bibfnamefont {L.}~\bibnamefont {Mahadevan}},\ }\href@noop {} {\bibfield
  {journal} {\bibinfo  {journal} {eLife}\ }\textbf {\bibinfo {volume} {7}},\
  \bibinfo {pages} {e34695} (\bibinfo {year} {2018})}\BibitemShut {NoStop}%
\bibitem [{\citenamefont {Cross}(2019)}]{cross2019}%
  \BibitemOpen
  \bibfield  {author} {\bibinfo {author} {\bibfnamefont {R.~A.}\ \bibnamefont
  {Cross}},\ }\href@noop {} {\bibfield  {journal} {\bibinfo  {journal} {Curr.
  Opin. Cell Biol.}\ }\textbf {\bibinfo {volume} {56}},\ \bibinfo {pages} {88 }
  (\bibinfo {year} {2019})}\BibitemShut {NoStop}%
\bibitem [{Note1()}]{Note1}%
  \BibitemOpen
  \bibinfo {note} {Provided we adjust the length of the shorter filament to
  account for the changing healing length $l_H \sim 1/n$, since the filaments
  are weakly bonded over this length scale.}\BibitemShut {Stop}%
\bibitem [{\citenamefont {Lake}\ and\ \citenamefont {Thomas}(1967)}]{lake1967}%
  \BibitemOpen
  \bibfield  {author} {\bibinfo {author} {\bibfnamefont {G.}~\bibnamefont
  {Lake}}\ and\ \bibinfo {author} {\bibfnamefont {A.}~\bibnamefont {Thomas}},\
  }\href@noop {} {\bibfield  {journal} {\bibinfo  {journal} {Proc. R. Soc.
  Lond. A}\ }\textbf {\bibinfo {volume} {300}},\ \bibinfo {pages} {108}
  (\bibinfo {year} {1967})}\BibitemShut {NoStop}%
\bibitem [{\citenamefont {Wang}\ \emph {et~al.}(2019)\citenamefont {Wang},
  \citenamefont {Panyukov}, \citenamefont {Rubinstein},\ and\ \citenamefont
  {Craig}}]{wang2019}%
  \BibitemOpen
  \bibfield  {author} {\bibinfo {author} {\bibfnamefont {S.}~\bibnamefont
  {Wang}}, \bibinfo {author} {\bibfnamefont {S.}~\bibnamefont {Panyukov}},
  \bibinfo {author} {\bibfnamefont {M.}~\bibnamefont {Rubinstein}}, \ and\
  \bibinfo {author} {\bibfnamefont {S.~L.}\ \bibnamefont {Craig}},\ }\href@noop
  {} {\bibfield  {journal} {\bibinfo  {journal} {Macromolecules}\ }\textbf
  {\bibinfo {volume} {52}},\ \bibinfo {pages} {2772} (\bibinfo {year}
  {2019})}\BibitemShut {NoStop}%
\bibitem [{\citenamefont {Hui}\ \emph {et~al.}(2003)\citenamefont {Hui},
  \citenamefont {Jagota}, \citenamefont {Bennison},\ and\ \citenamefont
  {Londono}}]{hui2003}%
  \BibitemOpen
  \bibfield  {author} {\bibinfo {author} {\bibfnamefont {C.-Y.}\ \bibnamefont
  {Hui}}, \bibinfo {author} {\bibfnamefont {A.}~\bibnamefont {Jagota}},
  \bibinfo {author} {\bibfnamefont {S.}~\bibnamefont {Bennison}}, \ and\
  \bibinfo {author} {\bibfnamefont {J.}~\bibnamefont {Londono}},\ }in\
  \href@noop {} {\emph {\bibinfo {booktitle} {Proc. R. Soc. Lond. A}}},\ Vol.\
  \bibinfo {volume} {459}\ (\bibinfo {year} {2003})\ pp.\ \bibinfo {pages}
  {1489--1516}\BibitemShut {NoStop}%
\bibitem [{\citenamefont {Zhao}(2014)}]{zhao2014}%
  \BibitemOpen
  \bibfield  {author} {\bibinfo {author} {\bibfnamefont {X.}~\bibnamefont
  {Zhao}},\ }\href@noop {} {\bibfield  {journal} {\bibinfo  {journal} {Soft
  Matter}\ }\textbf {\bibinfo {volume} {10}},\ \bibinfo {pages} {672} (\bibinfo
  {year} {2014})}\BibitemShut {NoStop}%
\bibitem [{\citenamefont {Creton}\ and\ \citenamefont
  {Ciccotti}(2016)}]{creton2016}%
  \BibitemOpen
  \bibfield  {author} {\bibinfo {author} {\bibfnamefont {C.}~\bibnamefont
  {Creton}}\ and\ \bibinfo {author} {\bibfnamefont {M.}~\bibnamefont
  {Ciccotti}},\ }\href@noop {} {\bibfield  {journal} {\bibinfo  {journal} {Rep.
  Prog. Phys.}\ }\textbf {\bibinfo {volume} {79}},\ \bibinfo {pages} {046601}
  (\bibinfo {year} {2016})}\BibitemShut {NoStop}%
\bibitem [{\citenamefont {Andrews}\ \emph {et~al.}(1987)\citenamefont
  {Andrews}, \citenamefont {Khan},\ and\ \citenamefont
  {Lockington}}]{andrews1987}%
  \BibitemOpen
  \bibfield  {author} {\bibinfo {author} {\bibfnamefont {E.}~\bibnamefont
  {Andrews}}, \bibinfo {author} {\bibfnamefont {T.}~\bibnamefont {Khan}}, \
  and\ \bibinfo {author} {\bibfnamefont {N.}~\bibnamefont {Lockington}},\
  }\href@noop {} {\bibfield  {journal} {\bibinfo  {journal} {J. Mater. Sci.}\
  }\textbf {\bibinfo {volume} {22}},\ \bibinfo {pages} {2833} (\bibinfo {year}
  {1987})}\BibitemShut {NoStop}%
\bibitem [{\citenamefont {Brown}(2007)}]{brown2007}%
  \BibitemOpen
  \bibfield  {author} {\bibinfo {author} {\bibfnamefont {H.~R.}\ \bibnamefont
  {Brown}},\ }\href@noop {} {\bibfield  {journal} {\bibinfo  {journal}
  {Macromolecules}\ }\textbf {\bibinfo {volume} {40}},\ \bibinfo {pages} {3815}
  (\bibinfo {year} {2007})}\BibitemShut {NoStop}%
\bibitem [{\citenamefont {Persson}(2013)}]{persson2013}%
  \BibitemOpen
  \bibfield  {author} {\bibinfo {author} {\bibfnamefont {B.~N.}\ \bibnamefont
  {Persson}},\ }\href@noop {} {\emph {\bibinfo {title} {Sliding friction:
  physical principles and applications}}}\ (\bibinfo  {publisher} {Springer
  Science \& Business Media},\ \bibinfo {year} {2013})\BibitemShut {NoStop}%
\end{thebibliography}%

\end{document}